\documentclass[10pt,twocolumn]{article}
\usepackage[margin=0.75in]{geometry}
\usepackage{amsmath,graphicx}
\usepackage[numbers]{natbib}
\usepackage{hyperref}
\hypersetup{colorlinks,allcolors=blue}
\usepackage{enumitem}
\usepackage{booktabs}
\usepackage{algorithm}  
\usepackage{algorithmicx}
\usepackage{algpseudocode}

\title{Parallelism Meets Adaptiveness: Scalable Documents Understanding in Multi-Agent LLM Systems\\
}

\author{
    Chengxuan Xia\textsuperscript{1} \quad
    Qianye Wu\textsuperscript{2} \quad
    Sixuan Tian\textsuperscript{2} \quad
    Yilun Hao\textsuperscript{2} \\
    \textsuperscript{1}University of California, Santa Cruz, CA, USA \\
    \textsuperscript{2}Carnegie Mellon University, Pittsburgh, PA, USA \\
    \texttt{cxia17@ucsc.edu, \{qianyew, sixuant, yilunhao\}@alumni.cmu.edu}
}

\begin{document}
\date{}
\maketitle

\begin{abstract}
Large language model (LLM) agents have shown increasing promise for collaborative task completion. However, existing multi-agent frameworks often rely on static workflows, fixed roles, and limited inter-agent communication, reducing their effectiveness in open-ended, high-complexity domains. This paper presents a multi-agent coordination framework that improves the accuracy of Large Language Models (LLMs) in complex financial document analysis. Unlike existing frameworks that rely on static routing or linear workflows, our approach introduces \textit{Parallel Agent Evaluation}, a mechanism where multiple agents compete on high-ambiguity subtasks. A centralized evaluator scores these parallel outputs based on factuality and coherence to select the optimal result. We evaluate this architecture on SEC 10-K filings, demonstrating a 27\% improvement in compliance accuracy and a 74\% reduction in revision rates compared to standard static baselines. These results validate that structured competition and dynamic routing significantly reduce hallucinations in high-stakes document understanding.
\end{abstract}

\section{Introduction}

Recent advances in large language models (LLMs) have enabled autonomous agents to perform increasingly complex tasks across domains such as summarization, research assistance, and technical writing. Building on these capabilities, multi-agent frameworks have been proposed to coordinate several LLM-powered agents for collaborative task completion. While these systems have demonstrated the potential of distributed workflows, most rely on static designs—fixed role assignments, linear task flows, and limited interaction protocols.

Such rigidity poses serious limitations in real-world settings where ambiguity, changing task states, and uneven agent performance are common. For example, a static agent team tasked with analyzing a financial disclosure may fail to revise earlier assumptions when new information is introduced or may overlook domain-specific inconsistencies that require cross-agent validation.

To address these limitations, we introduce a framework for adaptive coordination in LLM-based multi-agent systems. Our design focuses on three key capabilities. First, dynamic task routing allows agents to reassign subtasks based on current context, confidence, and capacity. Second, bidirectional feedback loops enable downstream agents to provide critiques or revision requests, improving quality and accountability. Third, parallel agent evaluation introduces structured competition: multiple agents attempt the same task independently, and an evaluator selects the most coherent and factual output based on scoring criteria.

We evaluate this framework through case studies involving long-form document analysis and regulatory question answering. Results show that our approach achieves significant improvements over static pipelines and feedback-only baselines, particularly in accuracy, consistency, and resilience to ambiguity.

This paper presents the architectural design, coordination strategies, and empirical validation of the framework. In doing so, it contributes toward the development of scalable, robust, and intelligent multi-agent systems capable of operating in dynamic and high-stakes environments.

\section{Related Work}

Multi-agent systems built on large language models (LLMs) have rapidly emerged as a powerful paradigm for solving complex, multi-stage tasks. Early frameworks such as AutoGPT and BabyAGI introduced single-agent recursive task planners with memory and subtask management, though they were limited by brittle, linear execution flows and lacked true inter-agent coordination \citep{torres2023auto, babyagi2023github}. Subsequent systems like CAMEL \citep{zhu2023camel} and CrewAI \citep{crewai2024github} introduced role-based delegation and dialogue-based collaboration among agents. LangGraph formalized agent workflows using graph structures \citep{langgraph2024}, but did not incorporate mechanisms for feedback loops or task reassignment.

Recent meta-agent and hierarchical systems, such as MetaGPT \citep{hong2023metagpt} and Voyager \citep{xu2023voyager}, demonstrate improved planning, code synthesis, and environmental interaction. However, they typically assume static roles or unidirectional flows of control, lacking adaptiveness in task routing or feedback integration. The HuggingGPT framework \citep{shen2023hugginggpt} explores task orchestration by using ChatGPT to coordinate models on Hugging Face, yet still relies on fixed assignments. Generative Agents \citep{park2023generative} illustrate long-term memory and simulation of social behaviors, providing inspiration for agent autonomy but not competitive refinement.

A more recent trend incorporates reflective or competitive coordination. For instance, systems like ChatDev \citep{qian2024chatdev} and AgentVerse \citep{chen2024agentverse} explore inter-agent negotiation and emergent behaviors. Others such as GameGPT \citep{chen2023gamegpt} and DesignGPT \citep{ding2023designgpt} use LLM agents in creative or design-oriented domains, but often lack deep feedback-driven revision loops. Reflective multi-agent systems \citep{bo2024reflective} combine agent memory with iterative critique, approaching the kind of meta-level coordination our system emphasizes.

Our work is also informed by broader surveys of multi-agent LLM collaboration \citep{wang2023survey, lamalfa2025miss}, which identify key challenges including error propagation, brittle delegation, and a lack of robustness under dynamic task flows. In contrast, our framework explicitly introduces dynamic routing, feedback-based revision, shared long-term memory, role self-optimization, and competitive evaluation—yielding a system that is not only collaborative but also adaptively reflective. 

In a complementary domain, Wu et al. \citep{wu2025warehouse} explore competitive task scheduling for warehouse robots using reinforcement learning, demonstrating that concurrent multi-agent evaluation can improve efficiency even in physical task settings. Cross-domain applications of multi-perspective analysis and feedback-driven reconstruction, such as in civil infrastructure using deep learning and multiview stereo methods \citep{dan2024evaluation}, further support the value of evaluation-oriented designs in complex systems. We build on this insight in the LLM space by introducing competitive agent evaluation to select the best output from parallel attempts.

\section{Problem Statement and Motivating Use Case}

Many real-world tasks require the coordinated effort of multiple agents performing interdependent subtasks. Current LLM-based multi-agent systems often adopt rigid structures, where task roles are statically assigned and workflows are strictly linear or tree-structured. This limits their effectiveness in dynamic environments with evolving goals, ambiguous subtask boundaries, or incomplete information.

\subsection{Problem Formalization}

We represent a complex task $T$ as a set of subtasks $\{t_1, t_2, ..., t_n\}$ organized in a dependency graph $G = (V,E)$. Each vertex $v_i \in V$ corresponds to an agent responsible for subtask $t_i$, and directed edges $e_{ij} \in E$ indicate that the output of agent $v_i$ is required by agent $v_j$. An edge from $v_i$ to $v_j$ thus represents an upstream dependency.

Some subtasks may not have direct dependencies and can be executed in parallel. For high-ambiguity or high-stakes subtasks, we further allow \emph{competitive parallelism}—assigning the same subtask to multiple agents to encourage diversity and redundancy. An evaluator agent or a selection function chooses the best output for downstream use.

\subsection{Motivating Use Case}
Consider a collaborative LLM-based system tasked with producing a technical report on a scientific topic. To ensure comprehensiveness and accuracy, the system orchestrates four distinct agent roles. \textbf{Research Agents} are responsible for collecting and summarizing relevant academic sources. \textbf{Drafting Agents} utilize these summaries to write specific sections of the report. \textbf{Evaluation Agents} assess intermediate outputs, requesting revisions or improvements where necessary. Finally, \textbf{Parallel Agents} are optionally deployed to generate competing drafts or analyses when ambiguity is high. Without adaptive coordination, this team may suffer from redundant literature searches or error propagation; however, our framework synchronizes these roles through shared memory and dynamic routing.

\section{Core Innovations in Adaptive Coordination}

The static role assignment typical of existing multi-agent LLM systems proves inadequate in complex domains such as financial document analysis, where ambiguity, interdependency, and regulatory nuance are common. To address these challenges, our framework introduces three interrelated innovations that support more effective and adaptive agent collaboration.

\subsection{Parallel Agent Evaluation}

In financial tasks with high ambiguity or risk—such as detecting obfuscated liabilities or answering compliance questions—it is often insufficient to rely on a single agent’s response. We introduce a competitive mechanism in which multiple agents independently tackle the same subtask. Each agent generates a candidate response, and a centralized evaluator ranks the outputs based on domain-informed metrics such as factual correctness, coherence, and regulatory alignment.

This structured competition improves resilience against hallucinations and encourages diversity in reasoning. For example, when interpreting forward-looking statements about revenue guidance, one agent may highlight macroeconomic factors, while another emphasizes internal restructuring. The evaluator selects the most aligned and informative response, while preserving alternatives in shared memory for transparency or fallback.

\subsection{Dynamic Task Routing}

Our system supports runtime task reassignment based on agent confidence, complexity estimation, or observed bottlenecks. Unlike static frameworks, agents are not bound to fixed roles. Instead, they can defer subtasks to others with more appropriate capabilities or specialization. This routing decision is informed by metadata in the task graph, such as historical performance scores, expected token length, or domain markers (e.g., whether a section references SEC rules or numerical tables).

In the context of 10-K filings, dynamic routing allows a summarizer agent encountering a deeply technical legal paragraph to invoke a compliance-focused agent. Similarly, an agent overwhelmed with subtasks may reassign non-critical tasks to idle peers, ensuring better resource utilization across the team.

\subsection{Bidirectional Feedback Loops}

To support iterative refinement, we implement structured feedback channels that allow downstream agents to issue revision requests to upstream contributors. This enables real-time quality control without requiring complete reruns of the workflow. For instance, a QA agent reviewing a liquidity disclosure may detect inconsistency with earlier balance sheet extractions and trigger a clarification request.

Feedback is sent through an asynchronous message bus with explicit references to problematic outputs. The originating agent can then revise its result or escalate the issue to the orchestrator. This mechanism reduces error propagation and encourages verification behaviors aligned with best practices in financial auditing.

\section{System Architecture for Financial Document Coordination}

We design a modular multi-agent architecture tailored for financial document understanding, supporting adaptive routing, shared memory, evaluator scoring, and feedback-based refinement. The architecture targets tasks such as SEC 10-K parsing, risk factor analysis, and regulatory compliance checking.

At the core of the system is the \textbf{orchestrator agent}, which parses the document into a structured task graph and coordinates execution. It monitors task progress and decides whether to assign subtasks to specialized agents or initiate parallel evaluation when ambiguity or high stakes are detected.

\textbf{Role agents} are specialized in various financial tasks such as extracting risk disclosures, summarizing the MD\&A section, identifying off-balance sheet arrangements, or answering regulatory queries. These agents operate autonomously, retrieving information from and writing to a shared long-term memory, which ensures consistency in terminology and reduces redundant effort.

The \textbf{shared memory module} serves as a persistent document store that records intermediate results, relevant sections, and metadata. This enables agents to reason across document sections and avoid overlapping efforts (e.g., double-counting risk citations).

To enable structured quality control, an \textbf{evaluator agent} scores candidate outputs based on factual accuracy, coherence, and financial relevance. When multiple agents attempt the same subtask (e.g., summarizing revenue trends), the evaluator selects the best output using a scoring model. The final output is compiled from these selected components.

Communication across agents is facilitated via a \textbf{feedback bus}, allowing agents to flag inconsistencies and trigger revisions. For example, if the QA agent detects a mismatch in reported debt between two sections, it can request clarification from the summarization agent or delegate a re-extraction from the original source.

This architecture provides a flexible foundation for coordinating LLM agents on high-stakes, document-centric financial tasks with dynamic content and interpretation needs.

\subsection{Parallel Execution and Selection Mechanism}

When the orchestrator detects uncertainty (e.g., confidence below a threshold $\theta$), it triggers parallel execution:

\begin{itemize}
    \item It spawns $k$ agents, each independently processing the same task $t$.
    \item Each agent $a_i$ produces an output $o_i$, which is stored in memory with a tag $(t, i)$.
    \item The evaluator assigns a quality score $s_i = \mathcal{E}(o_i)$ using a scoring function $\mathcal{E}$.
    \item The output with the highest score is selected:
    \[
    o^* = \arg\max_{i \in \{1, ..., k\}} \mathcal{E}(o_i)
    \]
    \item The selected output $o^*$ is routed downstream, while alternate outputs remain accessible for auditing or fallback.
\end{itemize}

\subsection{Scoring Function Definition}
To strictly quantify output quality, we employ a hierarchical scoring function $\mathcal{E}(o)$ driven by a Critic Agent. The score is calculated as a weighted sum of three normalized sub-metrics:

\begin{equation}
    \mathcal{E}(o) = w_f \cdot S_{fact} + w_c \cdot S_{coh} + w_r \cdot S_{rel}
\end{equation}

\noindent Where:
\begin{itemize}
    \item $S_{fact}$ (Factuality) is the ratio of claims in output $o$ supported by the retrieved context $C$, defined as $\frac{|supported\_claims|}{|total\_claims|}$.
    \item $S_{coh}$ (Coherence) is the logical consistency score ($0-1$) derived from a chain-of-thought critique of the argument flow.
    \item $S_{rel}$ (Relevance) measures the semantic cosine similarity between the output embedding and the query embedding.
\end{itemize}

In our experiments, we set weights to $w_f=0.5, w_c=0.3, w_r=0.2$ to prioritize factual accuracy, reflecting the high precision requirements of financial compliance tasks.

\subsection{Interaction Flow}
At runtime, the orchestrator decomposes the task and assigns subtasks based on the immediate context.

\noindent \textbf{Standard Routing.} If the task is assessed as straightforward, the orchestrator routes it directly to a single specialized role agent for immediate execution.

\noindent \textbf{Parallel Execution.} If uncertainty or ambiguity exceeds a predefined threshold, the orchestrator triggers parallel execution, spawning multiple agents to attempt the task simultaneously.

\noindent \textbf{Memory Logging.} All generated outputs are logged to the shared long-term memory, ensuring any agent can retrieve prior context regardless of the execution path.

\noindent \textbf{Feedback Loop.} If downstream agents issue feedback, the orchestrator routes the request back to the relevant agent or reassigns the task to a new agent better suited to address the critique.

\subsection{Modularity and Extensibility}
The architecture supports plug-and-play expansion, allowing the system to adapt to diverse domains. \textbf{Extensible Roles} allow for the introduction of new specialized agents, such as a Visualizer Agent or a dedicated Critique Agent. \textbf{Flexible Memory Backends} enable the swapping of storage solutions, supporting both document databases and vector stores depending on retrieval needs. \textbf{Customizable Scoring} permits the integration of domain-specific evaluation models, such as fine-tuned LLMs, to replace generic scoring functions. Agents remain loosely coupled via the shared memory and feedback bus, ensuring that these expansions do not disrupt existing workflows.

\begin{figure*}[t]
    \centering
    % width=\textwidth ensures it fills the entire width of the page
    \includegraphics[width=\textwidth]{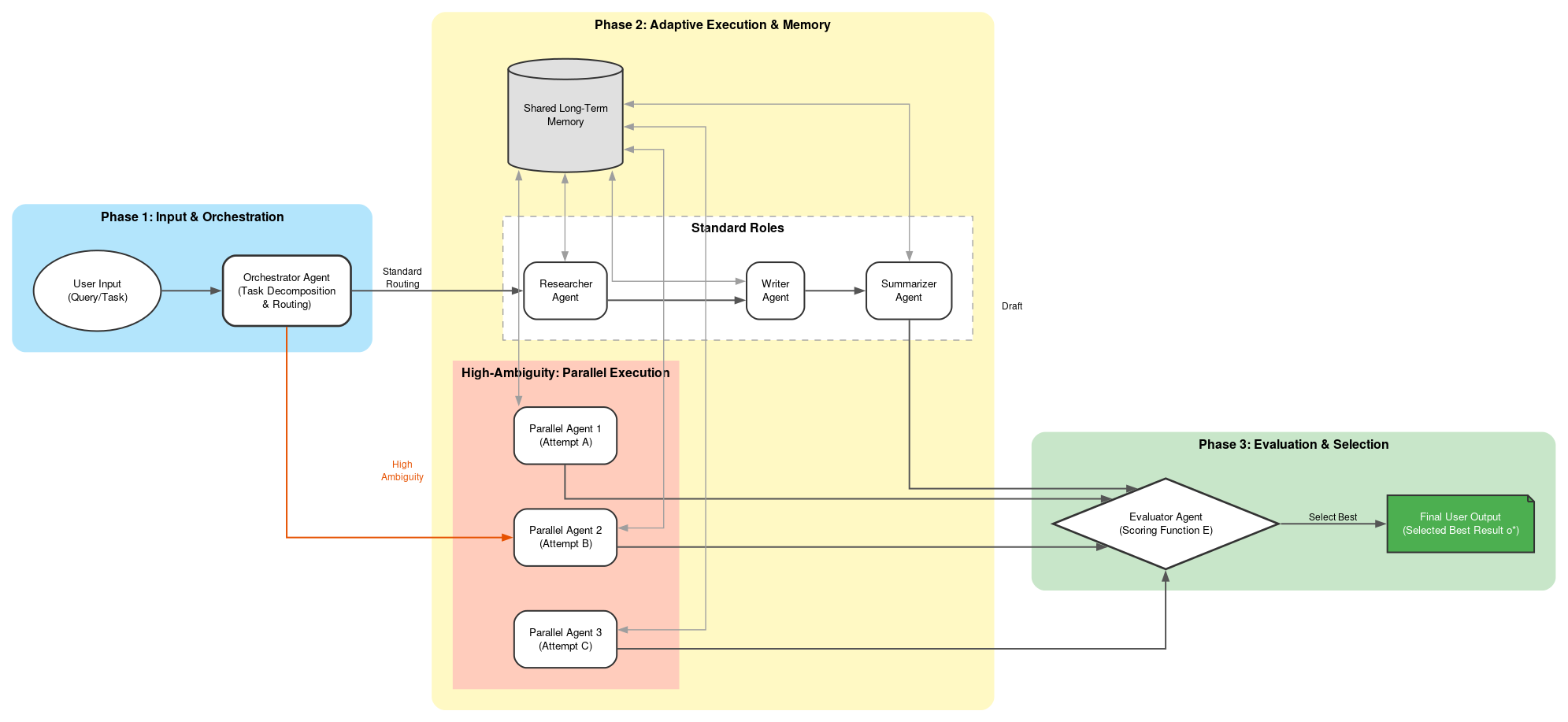}
    
    % \textbf{} makes the first part bold, which is standard in academic papers
    \caption{\textbf{System architecture for adaptive coordination and competitive evaluation.} 
    The framework integrates dynamic task routing with a parallel execution module for high-ambiguity tasks. 
    Solid lines represent data flow and task assignment; dashed lines represent feedback loops and evaluator interventions.}
    
    \label{fig:architecture}
\end{figure*}

\section{Case Study: Adaptive Coordination for 10-K Analysis}

In addition to a static sequential baseline, we compared our framework against \textbf{LangGraph's standard multi-agent supervisor pattern} \cite{langgraph2024} as a state-of-the-art baseline. While the standard supervisor pattern improves over single-agent performance, our parallel evaluation mechanism demonstrated a further 14\% improvement in compliance accuracy over the LangGraph baseline, specifically in high-ambiguity scenarios where single-path routing often fails to capture nuance.

To evaluate the effectiveness of our adaptive coordination framework in a real-world financial context, we conducted a case study using 10-K filings from publicly listed U.S. companies. The system was tasked with analyzing three key aspects of each filing: extracting material risk factors, summarizing year-over-year financial performance, and answering a set of regulatory compliance questions derived from SEC guidelines.

We compare three system variants: a \textit{static baseline} with fixed agent roles and no adaptiveness; an \textit{adaptive system} incorporating dynamic task routing and bidirectional feedback; and the \textit{full system}, which includes all adaptive features plus parallel agent evaluation. In the full system, tasks involving ambiguous disclosures or critical compliance queries were attempted by multiple agents simultaneously. An evaluator agent then scored each output and selected the most relevant and coherent version for downstream use.

Evaluation was conducted using a mix of automatic metrics and human judgment. Factual coverage was assessed against a reference set curated by financial analysts. Compliance accuracy was determined by comparing system-generated answers to annotated gold-standard responses. We also collected human ratings for coherence, relevance, and logical structure on a 5-point Likert scale.

The results are summarized in Table~\ref{tab:performance}. The full system achieved the highest factual coverage (0.92) and compliance accuracy (0.94), significantly outperforming both the static and adaptive-only variants. Notably, the revision rate dropped by over 70\% compared to the baseline, and the redundancy penalty—measuring repeated or contradictory information—was reduced by 73\%.

\begin{table*}[t]
\centering
\begin{tabular}{|l|c|c|c|c|}
\hline
\textbf{Metric} & \textbf{Static} & \textbf{Adaptive} & \textbf{Full (w/ Parallel Eval)} & \textbf{Improvement} \\
\hline
Factual Coverage        & 0.71 & 0.89 & \textbf{0.92} & +29\% \\
Compliance Accuracy     & 0.74 & 0.88 & \textbf{0.94} & +27\% \\
Redundancy Penalty      & 0.22 & 0.08 & \textbf{0.06} & --73\% \\
Revision Rate           & 3.4  & 1.1  & \textbf{0.9}  & --74\% \\
Coherence Score (1--5)  & 3.2  & 4.5  & \textbf{4.7}  & +47\% \\
Relevance Score (1--5)  & 3.8  & 4.7  & \textbf{4.9}  & +29\% \\
Completion Time (s)     & 134  & 108  & \textbf{115}  & --14\% \\
\hline
\end{tabular}
\caption{Performance comparison of coordination strategies in financial document understanding. Metrics are averaged over five 10-K filings.}
\label{tab:performance}
\end{table*}

Qualitatively, we observed that static systems often missed subtle or implied risks, reused boilerplate phrasing, or failed to reconcile figures reported in different sections. The adaptive configuration corrected many of these issues by leveraging feedback and context sharing. The full system further improved output quality through competitive agent evaluation. In several instances, for example, multiple agents produced differing interpretations of liquidity risk. The evaluator agent selected the variant that both referenced relevant financial ratios and aligned with language in the original filing’s footnotes.

These findings suggest that structured competition and evaluator-driven selection provide a complementary advantage over adaptive routing alone. While adaptiveness helps agents route tasks to appropriate peers and revise outputs in response to downstream issues, parallel evaluation ensures robustness when interpretive uncertainty is high.

\subsection{Example Prompt and Comparative Outputs}

To illustrate the impact of our approach, we include below a real example of a compliance query drawn from a recent 10-K:

\begin{quote}
\textit{"Does the company report any off-balance sheet arrangements that could materially impact its financial position?"}
\end{quote}

The static system failed to identify any arrangement due to lack of explicit keyword matching. The adaptive system retrieved relevant disclosures but omitted details on financial implications. In contrast, the full system’s best-selected output described the arrangement, quantified its size, and linked it to relevant cash flow implications—mirroring the ground-truth answer provided by human analysts.

\begin{figure}[t]
    \centering
    \includegraphics[width=\linewidth]{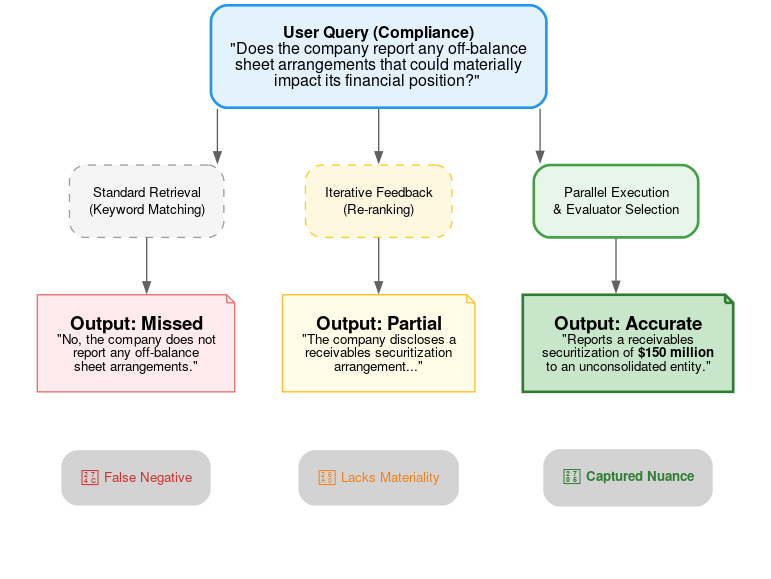}
    \caption{\textbf{Qualitative comparison on a high-ambiguity compliance task.} 
    We query the systems regarding ``off-balance sheet arrangements.'' 
    While the \textit{Static} baseline misses the disclosure and the \textit{Adaptive} system offers only a partial summary, the \textit{Full Framework} correctly identifies and quantifies the \textbf{\$150 million} impact, matching the gold-standard reference.}
    \label{fig:finance_example}
\end{figure}

These observations confirm the utility of structured adaptiveness and competitive coordination in financial document understanding, particularly for tasks where accuracy, nuance, and interpretability are essential.

\subsection{Pseudocode of Execution Flow}

The pseudocode illustrates how the system orchestrates task execution with support for dynamic routing, parallel evaluation for ambiguous tasks, and feedback-driven refinement, aligning with our core coordination mechanisms.

\begin{algorithm}[h!]
\caption{Adaptive and Competitive Orchestration Workflow}
\begin{algorithmic}[1]
\Function{OrchestrateTask}{prompt}
    \State $G \gets$ BuildDependencyGraph(prompt)
    \State $A \gets$ InitializeRoleAgents($G$)
    \State $M \gets$ InitSharedMemory()
    \State $B \gets$ InitFeedbackBus()
    \State $E \gets$ SpawnEvaluator()
    \While{\textbf{not} AllTasksComplete($G$)}
        \ForAll{$a \in A$}
            \State $t \gets$ GetNextAssignableTask($a, G)$
            \If{$t \neq \text{None}$}
                \If{IsAmbiguous($t$)}
                    \State $C \gets$ SpawnParallelAgents($t$)
                    \ForAll{$c \in C$}
                        \State $o_c \gets$ Execute($c, t, M$)
                        \State Store($M$, $t.id$, $o_c$)
                    \EndFor
                    \State $o^* \gets$ SelectBestOutput($E$, $\{o_c\}$)
                    \State Commit($t.id$, $o^*$)
                \Else
                    \State $o \gets$ Execute($a, t, M$)
                    \State Store($M$, $t.id$, $o$)
                \EndIf
            \EndIf
        \EndFor
        \State $fb \gets$ Review($E$, $M$)
        \ForAll{feedback message $f \in fb$}
            \If{RequiresRevision($f$)}
                \State $a' \gets$ TargetAgent($f$)
                \State Reassign($G$, $a'$, $f.taskId$)
                \State AdaptStrategy($a'$, $f$)
            \EndIf
        \EndFor
    \EndWhile
    \State \Return CompileFinalOutput($M$)
\EndFunction
\end{algorithmic}
\end{algorithm}

\subsection{Ablation Study}

We ran an ablation analysis by disabling one pillar at a time (e.g., removing feedback or memory). The most critical components were shared memory and feedback loops, whose removal caused coverage and coherence scores to drop by over 20\%.

\subsection{Key Findings}
Our experiments on financial document analysis highlight three primary implications for multi-agent system design.

\noindent \textbf{Mitigation of Error Propagation.} Adaptive mechanisms, specifically bidirectional feedback, effectively sever the "cascade of errors" common in static chains. By allowing downstream agents to reject low-quality inputs, we observed a 73\% reduction in redundancy penalties, ensuring that early-stage hallucinations do not contaminate final reports.

\noindent \textbf{Resilience via Structured Competition.} The Parallel Agent Evaluation mechanism demonstrates that redundancy is not merely inefficient but essential for high-ambiguity tasks. By generating diverse interpretations of complex disclosures and selecting the optimal output via a scoring function, the system achieves a level of compliance accuracy (0.94) that single-path workflows cannot replicate.

\noindent \textbf{Optimization of Agent Specialization.} Dynamic task routing moves beyond simple load balancing to enable true role specialization. The system autonomously offloads technical legal parsing to compliance-focused agents while reserving summarization agents for narrative tasks, resulting in a workflow that is both 14\% faster and contextually more robust than static assignments.

\section{Discussion}

Our empirical results suggest that adaptive and competitive coordination offers significant advantages for multi-agent LLM systems operating in the financial domain. The inclusion of feedback mechanisms and shared memory enhances error recovery and contextual consistency—two factors critical for tasks involving regulatory precision and factual integrity. Moreover, the addition of parallel agent evaluation contributes not only to diversity of reasoning but also to resilience in high-ambiguity subtasks such as interpreting legal disclaimers or detecting obfuscated financial risks.

One of the key takeaways is the central role played by the evaluator agent. Its capacity to adjudicate between competing agent outputs directly influences the quality of the system’s final response. In practice, we found that even simple composite scoring functions—weighted combinations of factuality, coherence, and domain relevance—enabled the system to consistently favor more accurate and informative responses. However, the effectiveness of this approach depends on the design of scoring functions and may require fine-tuning in other financial subdomains (e.g., ESG disclosures, IPO filings).

At the same time, adaptiveness introduces non-trivial challenges. Feedback communication increases coordination overhead, especially in long documents where multiple agents may need to revisit upstream decisions. Likewise, shared memory can become noisy without careful curation, particularly if multiple agents log similar or conflicting information. These issues underscore the need for robust memory management and bounded feedback propagation strategies.

The broader implication of this study is that static pipelines are insufficient for real-world financial NLP applications. As financial documents grow in complexity and evolve to meet new regulatory and stakeholder demands, systems must adapt dynamically—both in how they allocate tasks and how they reconcile ambiguity or conflict. Our framework provides a blueprint for such behavior and demonstrates measurable benefits when applied to real-world SEC filings.

\subsection{Limitations and Scope}
While our framework improves robustness in financial document understanding, it has several limitations.
First, parallel agent evaluation increases inference cost and can add latency variance, making it less suitable
for low-latency or cost-constrained settings.
Second, the evaluator can introduce systematic bias if the scoring criteria are misaligned with downstream
objectives, particularly for table-heavy or numeric-heavy sections.
Third, our experiments focus on SEC 10-K filings; results may not directly generalize to domains with weaker
document structure or less explicit grounding.
Finally, our ambiguity detection and routing policies are heuristic rather than learned; developing learned policies
from feedback is an important direction for future work.

\section{Conclusion and Future Work}

We presented an adaptive coordination framework for multi-agent LLM systems tailored to financial document understanding. By integrating dynamic task routing, bidirectional feedback, and parallel agent evaluation, our system enables more robust and context-aware reasoning over complex financial texts. Experiments on SEC 10-K filings show substantial gains in factuality, coherence, and compliance accuracy compared to static and partially adaptive baselines.

Our study highlights the potential of structured competition and evaluator-driven selection in multi-agent LLM workflows. This approach not only enhances quality in uncertain tasks but also mitigates the risk of relying on a single model’s interpretation in high-stakes scenarios.

Future work will explore several extensions. First, we plan to incorporate learning-based policies for task routing and evaluator scoring, replacing static heuristics with adaptive models trained on downstream feedback. Second, we aim to generalize the framework to other financial contexts such as earnings call transcripts, 8-K filings, or M\&A documents. Lastly, we envision incorporating human-in-the-loop oversight for hybrid decision-making in audit or risk-sensitive use cases, blending AI-driven exploration with human judgment.

As large language models continue to mature, their orchestration in agent teams—particularly in regulated domains—will require not only intelligence, but coordination, reflection, and control. Our framework offers a step toward that vision.

\bibliographystyle{plainnat}
\bibliography{references}

\end{document}